\documentclass[runningheads]{llncs}

\usepackage{times}
\usepackage{latexsym}
\usepackage{amsmath}
\usepackage{bbm}
\usepackage{calrsfs}
\usepackage{graphics}
\usepackage{algorithm2e}
\usepackage{algpseudocode}
\usepackage[normalem]{ulem}
\usepackage{rotating}

\usepackage{graphicx}
\usepackage{todonotes}
\usepackage[section]{placeins}

\begin{document}
\title{Ising-Based Louvain Method:\\Clustering Large Graphs with Specialized Hardware} 
%
%

\author{Pouya Rezazadeh Kalehbasti\thanks{Work done while at Fujitsu Laboratories of America, Inc}\inst{1} \and 
Hayato Ushijima-Mwesigwa\inst{2} \and 
Avradip Mandal\inst{2}  \and
Indradeep Ghosh\inst{2}
} 
\authorrunning{P. R.Kalehbasti et al.}
%
\institute{School of Engineering, Stanford University, Stanford CA 94305 \\ \email{pouyar@stanford.edu} \and
Fujitsu Laboratories of America, Inc., Sunnyvale CA 94085\\ \email{\{hayato,amandal,ighosh\}@fujitsu.com}}
\maketitle              
%

\begin{abstract}

Recent advances in specialized hardware for solving optimization problems such quantum computers, quantum annealers, and CMOS annealers give rise to new ways for solving real-word complex problems. However, given current and near-term hardware limitations, the number of variables required to express a large real-world problem easily exceeds the hardware capabilities, thus hybrid methods are usually developed in order to utilize the hardware. In this work, we advocate for the development of hybrid methods that are built on top of the frameworks of existing state-of-art heuristics, thereby improving these methods. We demonstrate this by building on the so called Louvain method, which is one of the most popular algorithms for the Community detection problem and develop and Ising-based Louvain method. The proposed method outperforms two state-of-the-art community detection algorithms in clustering several small to large-scale graphs. The results show promise in adapting the same optimization approach to other unsupervised learning heuristics to improve their performance.


\keywords{Graphs \and Community detection \and Clustering \and Ising Model}

\end{abstract}

\section{Introduction}
Graphs are powerful tools for modeling and analyzing complex systems from social networks to biological structures. In graphs, some groups of nodes can show modular behavior \cite{lancichinetti2008benchmark} in that nodes in these groups are highly interrelated and show similar traits and strong connections \cite{fortunato2010community}.
Communities or clusters are a form of these modules in graphs which are identified as clusters of nodes with a higher density of internal edges compared to external edges \cite{newman2016community,girvan2002community}. These highly inter-connected nodes reveal underlying structures within a network ranging from friend groups in social networks \cite{newman2004detecting}, similar proteins in biochemical structures \cite{krogan2006global}, and papers of similar categories in citation graphs \cite{ruan2013efficient}. \emph{Community Detection} (CD) is an unsupervised learning method which aims to identify these interwoven structures (communities, aka clusters) in graphs. CD is an NP-Hard combinatorial optimization problem.

In recent years, we have seen the emergence of specialized hardware designed for solving combinatorial optimization problem. Examples of these novel computing hardware include, adiabatic quantum computers, CMOS annealers, memristive circuits, and optical parametric oscillators, that are designed to solve optimization problems formulated as an Ising or QUBO mathematical model. Given that many well-known problems in graphs can easily be model in this form, there has been a growing interest in formulating and evaluating these problem and their subsequent QUBO models on the different specialized hardware platforms \cite{kumar2018quantum,shaydulin2019network,ushijima2017graph,cohen2020ising,negre2020detecting}. However, many real-world problems require significantly more variables than these devices can handle, thus hybrid methods are usually used.

\paragraph{Our contribution:}
This paper introduces a hybrid method, inspired by the Louvain algorithm, for solving the community detection problem on specialized hardware. Following the work on leveraging specialized hardware for local search \cite{liu2019modeling}, the proposed method he proposed method is expected to produce more optimal clusterings (\textit{i.e.} communities) compared to the original algorithm. This improvement is expected since instead of using a greedy approach for assigning nodes to communities iteratively, this method creates local optimization problems for a set of multiple nodes and multiple candidate clusters concurrently. The method is used to cluster several benchmark graphs where the results will show improved modularity scores compared to Louvain algorithm across several runs.

\subsection{Related Work}
Researchers have developed several methods for finding high-quality communities that can reveal the interworking and internal structures of complex networks. Increasing availability of large-scale graphs, like the World Wide Web, or social networks of Twitter, Facebook, and Instagram with millions or  billions of nodes and edges, have made traditional approaches for community detection unwieldy in terms of computational and spatial complexity \cite{blondel2008fast}. Consequently, researchers have devised efficient algorithms, $e.g.$ Girvan-Newman \cite{girvan2002community}, Louvain \cite{blondel2008fast}, and Leiden \cite{traag2019louvain} algorithms, for effectively detecting communities within large graphs. Some of the other well-known CD algorithms listed by Javed et al. \cite{javed2018community} include Hierarchical Clustering \cite{johnson1967hierarchical}, Graph Partitioning \cite{newman2013spectral}, Spectral Optimization \cite{newman2004finding}, Simulated Annealing \cite{guimera2005functional}, Clique Percolation Method (CPM) \cite{palla2005uncovering}, and Fuzzy Detection \cite{reichardt2004detecting}. Many of these algorithm use \emph{modularity} as a metric for solving the optimization problem of CD: such algorithms try to assign nodes to clusters which can maximize the modularity of the entire graph \cite{newman2004finding,newman2013spectral}. 

\section{Methods}
\subsection{Modularity Maximization}
The objective function selected here for clustering is Modularity which was introduced by Newman \cite{newman2004finding}. 
This metric favors highly interconnected nodes to be clustered together. For a graph, $G=(V,E)$, with $|V|$ the set of nodes and edge-set $E$, the Modularity Maximization problem maximizes the following objective function: 

\begin{equation} \label{eq:modularity}
    Q = \frac{1}{2m} \sum_{i,j} (A_{i,j} - \frac{k_i k_j}{2m}) \delta(c_i,c_j)
\end{equation} 

\noindent here, $A_{i,j}$ is the $(i,j)$ element in the adjacency matrix of $G$. $k_i$ is the weighted degree of node $i \in V$, and $m = |E|$. In addition, $c_i$ is the community node $i$ belongs to, and $\delta(u,v)$ equals 1 if $u=v$, and equals 0 otherwise. 


\subsection{QUBO Models}
Quadratic Unconstrained Binary Optimization (QUBO) models include 0/1 binary variables $q_i$, biases $c_i$, and couplers $c_{i,j}$. The objective is to minimize the following: 
\begin{equation} 
 E(q_1, \dots, q_n) = \sum_{i = 1} ^ n c_iq_i + \sum_{i<j} c_{i,j}q_{i}q_{j}
  \end{equation}
  QUBO models can be converted to Ising models (variables are $\pm 1$) and vice versa \cite{bian2010ising}.
  The objective function of the Modularity Maximization for at most 2-clusters is naturally defined in the Ising Model \cite{negre2020detecting}. In addition, the problem of at most $k$ clusters forms a quadratic function, thus leads to a natural formulation as a QUBO. Subsequently, there has been a large body of research utilizing specialized hardware for tackling this problem ranging from both small to large scale graphs \cite{ushijima2017graph,shaydulin2018community,shaydulin2019network,negre2020detecting,ushijima2019multilevel}.
\subsection{Solving Community Detection on the Fujitsu Digital Annealer}
The Fujitsu Digital Annealer (DA) is a
specialized architecture for combinatorial optimization problems formulated as QUBO \cite{aramon2019physics,daweb}. 
We use the second generation of the DA that is capable of solving problems with up to 8192 variables and up to 64 bits of precision. The DA has previously been used 
in different areas such as communication \cite{naghsh2019digitally}, signal processing \cite{rahman2019ising}, and data mining \cite{cohen2020ising,cohen2020ising2}. 

\subsection{Baseline Method: Louvain Algorithm}
Louvain Algorithm (aka Louvain) is a hierarchical clustering method that uses, mainly, modularity for clustering large graphs \cite{blondel2008fast,newman2016equivalence}. Louvain identifies the optimal clusters for nodes in an agglomerative hierarchical approach: it first assigns each node to an individual cluster, then identifies the optimal cluster for each node in its immediate neighboring clusters, and after several passes over the graph, it aggregates the nodes belonging to each cluster into super-nodes  \cite{blondel2008fast}. The same process is applied to these super-nodes until all clusters become singletons and no nodes can be assigned to clusters other than their own. The resulting super-nodes are the ultimate clusters for their associated sub-nodes. Blondel et al. \cite{blondel2008fast} and Traag et al. \cite{traag2019louvain} provide detailed illustrations of the Louvain algorithm. Next section introduces the Ising-Louvain where an optimization-based approach replaces the greedy approach in the original Louvain.

\subsection{Ising-Louvain Method}
The proposed method in this paper, Ising-Louvain, is an agglomerative hierarchical clustering method. This method can be summarized in the following steps
\begin{enumerate}
    \item Initially, each node is assigned to a separate cluster
    \item A breadth first Search is performed for each node to find its neighboring nodes, and selects a set of candidate clusters for each of these selected nodes
    \item These candidate clusters get assigned to the selected nodes in a way that the new assignments maximize the modularity of the entire graph
    \item Steps 2 and 3 are repeated until no more improvement in the modularity of the graph is possible
    \item Afterwards, the nodes belonging to the same cluster get aggregated into super-nodes, and steps 2-4 are applied to these super-nodes
    \item Steps 2-5 get repeated until all clusters become singletons.
\end{enumerate}

\noindent In the steps above, Step 3 is where the QUBOs are formed and solved. Specialized hardware, \textit{e.g.} Fujitsu's Digital Annealer, can effectively solve such optimization problems and yield high-quality solutions.

\begin{figure}
    \centering
    \includegraphics[width=0.6\textwidth]{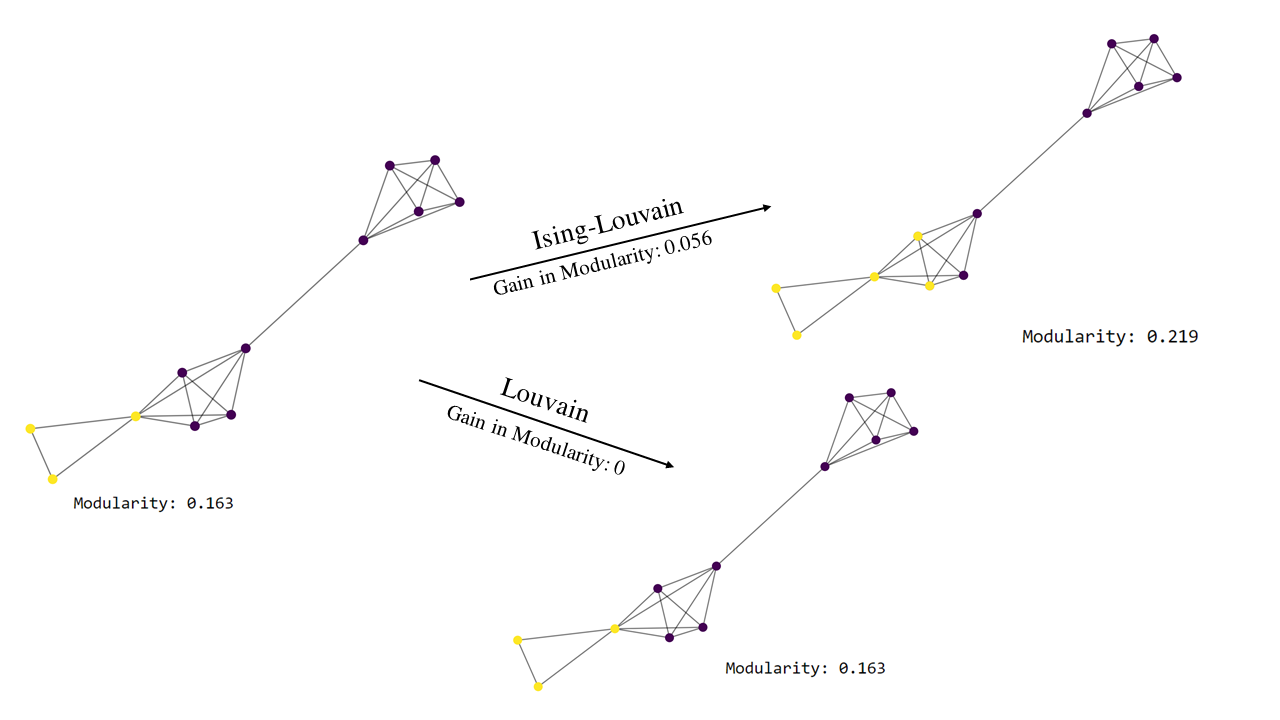} \vspace*{-5pt}
    \caption{Sample case where only Ising-Louvain can improve the graph while Louvain stalls}
    \label{fig:2-2} \vspace*{-15pt}
\end{figure}

Figure \ref{fig:2-2} shows a case where Louvain cannot improve the clustering of a graph in its current situation, while Ising-Louvain can. Here, Ising-Louvain can find a more optimal clustering by considering two nodes [or more than one node in general] at a time and changing the clusters of both nodes from violet to yellow. However, Louvain only considers a single node at each step and finds no gain in modularity in assigning a new cluster to that single node, and hence can never achieve the more optimal clustering that Ising-Louvain can obtain. 

\subsection{Formulating Ising-Louvain}

\subsubsection{Deriving the Objective Function}


What follows formulates the modularity as the optimization objective for the QUBO used in Ising-Louvain algorithm. Assume we have a graph with $n$ nodes and an edge set $E$. We start by selecting a set $S$ of nodes as our `free nodes' and by checking if they should belong to a specific community $C$ or not. The community-membership of the rest of the nodes are assumed to be constant. Let $\boldsymbol{x} \in \{0,1\}$ be a vector of $n$ binary variables indicating whether or not each node in the graph belongs to community $C$, such that $x_i = 0$ if $i \in C$ and $x_i = 0$ otherwise. Following a similar derivation as Negre et al. \cite{negre2020detecting}, the objective function defined in Equation \ref{eq:modularity} can be written as follows: \vspace*{-5pt}

\begin{equation} \label{eq:26}
         \max(Q) \equiv \min_{\boldsymbol{x}} \Big( \sum_{i \in S} \sum_{j \in S} x_i (\frac{k_i k_j}{2m} - A_{i,j}) x_j + \sum_{i \in S} x_i \big( 2 \sum_{j \in C} (\frac{k_i k_j}{2m} - A_{i,j}) \big) \Big)
\end{equation}

\noindent We can apply the same process to nodes in set $S_l$ and any other cluster $C_l$ and obtain similar phrases to Equation \ref{eq:26}. When considering more than one cluster in our optimization problem, we should consider the constraint that each node in $S$ (defined as $\bigcup_l S_l$) should belong to one and only one cluster at each instant. Equation \ref{eq:27} shows the objective function resulting from superposing the objective functions for a set of candidate clusters, called set $L$, with a penalty term imposing the mentioned constraint:
\begin{multline}
\begin{aligned}
         \min_{\boldsymbol{x}} \bigg( &\sum_{l \in L} \Big( \sum_{i \in S_l} \sum_{j \in S_l} x_{i,l} (\frac{k_i k_j}{2m} - A_{i,j}) x_{j,l} + \sum_{i \in S_l} x_{i,l} \big( 2 \sum_{j \in C_l} (\frac{k_i k_j}{2m} - A_{i,j}) \big) \Big)
         \\&+ \gamma \sum_{i \in S_l} (\sum_{l \in L} x_{i,l} - 1)^2 \bigg)
\end{aligned} \label{eq:27}
\end{multline}
\noindent where $l$ indicates the cluster number, $S_l$ indicates the nodes for which cluster $C_l$ is a candidate, $x_{i,l}$ is the i'th element of vector $x_l$ which contains the binary variables indicating the membership of all nodes to cluster $C_l$. The last term in this equation imposes the constraint mentioned above with $\gamma$ being the penalty coefficient.

Equation \ref{eq:27} 
has a space complexity 
of $O\big(\max{\{|S|^2, \mathcal{L}\}}\big)$ where $\mathcal{L} = \sum_l |S_l|$. The optimization objective has $\mathcal{L}$ variables
, so the QUBO is of size $\mathcal{L} \times \mathcal{L}$. 

\subsubsection{Pseudocode}
Algorithm \ref{alg:1} shows the pseudocode of the proposed method [partially following the pseudocodes of Blondel et al. \cite{blondel2008fast} and Traag et al. \cite{traag2019louvain}]. Table \ref{tab:functions_and_vars} describes functions and auxiliary variables used in Algorithm \ref{alg:1}, and Table \ref{tab:hyperparams} describes the hyper-parameters defined to fine-tune the behavior of the algorithm. 

\begin{table}[!htbp]
    \centering
        \caption{Definition of functions and auxilliary variables used in Algorithm \ref{alg:1}}
    \begin{tabular}{l|l}
        \textbf{Function/Variable Name} & \textbf{Definition}\\
        \hline
        Singleton(G) & Create singleton clusters for graph G \\
        Aggregate(G, P) & Aggregate nodes in the same clusters (based on partitioning P) in\\
        &\ graph G into supernodes\\
        modularity(G, P) & Modularity of graph G with partitioning P\\
        ModGain(P, X) & Modularity gain if partition P is updated with assignments X\\
        modified & flag indicating if graph is modified throughout the inner loop\\
        done & flag indicating that graph can't be improved anymore, \textit{i.e.} all clusters\\
        &\ have stayed as singletons after running the inner while-loop\\
        $nodes$ & list of nodes in graph G
    \end{tabular} 
    \label{tab:functions_and_vars} \vspace*{-40pt}
\end{table}


\begin{table}[!htbp]
    \centering
        \caption{Definition of hyper-parameters used in Algorithm \ref{alg:1}}
    \begin{tabular}{l|l}
        \textbf{Hyperparameter Name} & \textbf{Definition}\\
        \hline
         Max\_Nodes &  Maximum number of nodes that the algorithm selects as set $S$\\
         Max\_Clusters & Maximum number of candidate clusters that the algorithm considers\\
         &\ for each node in set $S$\\
         Max\_Node\_Visits & Maximum number of times the algorithm includes each node of the\\
         &\ graph in set $S$ over a single pass of the graph during the inner while-loop\\
         random\_seed & Seed number used to initialize the random generator in Numpy\\
         solver\_timeout & Timeout limit for each call to the solver\\
         BFS\_Depth & Depth of the BFS run by the algorithm (explained more in Section \ref{strategies})\\
         gamma & the penalty coefficient in QUBO\\
         $counter\_max_{out}$ & maximum allowed number of iterations for the outer while-loop\\
         $counter\_max_{in}$ & maximum allowed number of iterations for the inner while-loop\\
         $\theta$ & minimum acceptable improvement in modularity over an inner or outer\\
         &\ while-loop \\ \hline
    \end{tabular}
    \label{tab:hyperparams} 
\end{table}


\begin{algorithm}
\SetAlgoLined
\SetKwFunction{FMain}{IsingLouvain}
\SetKwProg{Fn}{Function}{:}{}
\Fn{\FMain{Graph G}}{
    P = Singleton(G)\;
    P = RefineAndCoarsen(G, P)\;
    \Return P
    }
\vspace{5pt}


\SetKwFunction{FF}{RefineAndCoarsen}
\SetKwProg{Fn}{Function}{:}{}
\Fn{\FF{Graph G, Partition P}}{
    $counter_{out} = 0$\;
    $\Delta Q_{out} = \theta$\;
    $done = False$\;
    \tcc{Outer While Loop}
    \While{\big(($done == False$) \& ($counter_{out} \leq counter\_max_{out}$) \& ($\Delta Q_{out} \geq \theta$)\big)}{
        $counter_{out} \mathrel{+}= 1$\; 
        $counter_{in} = 0$\;
        $Q_{old,out} = modularity(G, P)$\;
        $\Delta Q_{in} = \theta$\;
        $modified = True$\;
        \tcc{Inner While Loop}
        \While{\big(($modified == True$) \& ($counter \leq counter\_max$) \& ($\Delta Q_{in} \geq \theta$)\big)}{
            $counter_{in} \mathrel{+}= 1$\; 
            $modified = False$\;
            $Q_{old,in} = modularity(G, P)$\;
            \tcc{Doing a Single Pass}
            P, modified = RunOnePass(G, P, modified)\;
            $Q_{new,in} = modularity(G, P)$\;
            $\Delta Q_{in} = Q_{new,in} - Q_{old,in}$\;
            $Q_{old,in} = Q_{new,in}$\;
            }
        $Q_{new,out} = modularity(G, P)$\;
        $\Delta Q_{out} = Q_{new,out} - Q_{old, out}$\;
        $Q_{old,out} = Q_{new,out}$\;
        \eIf{$|P| == n$}{
                done $\leftarrow$ True\;
            }{
                G = Aggregate(G, P)\;
                P = Singleton(G)\;
            }
        }
    \Return G, P
    }
\vspace{5pt}


\SetKwFunction{FFF}{RunOnePass}
\SetKwProg{Fn}{Function}{:}{}
\Fn{\FFF{Graph G, Partition P, Bool mod}}{
    \For{$i \in nodes$}{
        Select $S$ and $L$\;
        Calculate $\boldsymbol{\mathcal{B}}_{\mathcal{L} \times \mathcal{L}}, \boldsymbol{\mathcal{B}}_{\mathcal{C} \times \mathcal{L}}, \boldsymbol{\mathcal{Z}}, \boldsymbol{\zeta}^T$\;
        $\boldsymbol{X} = [\boldsymbol{x}_{l,S_l}]_{\mathcal{L} \times 1}$ (\textit{i.e.} solution to Equation \ref{eq:27})\; 
        \If{$ModGain(P, X) > 0$}{
            Update P with $X$\;
            mod $\leftarrow$ True\;
            }
        }
    \Return P, mod
    }

\caption{Ising-Louvain} 
\label{alg:1}
\end{algorithm} 

\subsection{Implementation Details of Ising-Louvain}
The first part of this section will study the strategies tried for selecting candidate nodes and clusters and the ones which yielded the best results, the second part studies the rule-of-thumb devised for tuning the $\gamma$ hyperparameter in Equation (\ref{eq:27}). Lastly, we summarise the different approaches taken to reduce the time complexity of the proposed method. 

\subsubsection{Node and Cluster Selection Strategies} \label{strategies}
The authors noticed that the strategy for selecting nodes and candidate clusters has a sizable impact on the run-time and performance (measured by the obtained optimal modularity value) of the devised clustering algorithm. The explored node-selection-strategies for forming set $S$ were as follows:
\begin{enumerate}
    \item Random strategy: selecting set $S$ at random from the graph
    \item Sliding-window strategy: selecting set $S$ from consecutive nodes in the sorted list of nodes, and shifting the list by a random number after each pass over the entire graph [shifting would result in (probably) different selections during each pass]
    \item BFS strategy: selecting set $S$ based on a breadth-first-search (BFS), of specified depth, originating from each node selected sequentially from the shuffled list of nodes
\end{enumerate}
Experiments showed that the third strategy yielded consistently better results across different graphs and across different runs on the same graphs. The experiments reported later on are conducted with this approach for node selection.

The following strategies, for selecting the cluster-set $L$, were considered in this work.  
\begin{enumerate}
    \item BFS strategy: selecting set $L$ at random from clusters resulting from a BFS [of specified depth] originating in each node, while assuring the current clusters for nodes in set $S$ are included in the candidate clusters for each node [this extra step enabled obtaining a trivial solution to the QUBO with all nodes kept in their current clusters]
    \item Semi-greedy strategy: composing set $L$ from the top $k$ clusters for each node in set $S$ based on a greedy search for that node [the greedy search would inspect all neighboring clusters as well as the original cluster for the node, and would select the top $k$ clusters which resulted in the highest gains in modularity if the node was to be moved to those clusters]
\end{enumerate}
The second strategy resulted in consistently better results across different graphs and across different runs on the same graphs, so this strategy was used to conduct the experiments reported later in the paper. 

\subsubsection{Strategy for Tuning $\gamma$}
The authors devised the following rule-of-thumb for setting the initial value of $\gamma$ in Equation \ref{eq:27}:
\begin{equation*}
         \gamma = max(k_i)\ \text{for}\ i \in G
\end{equation*}
This initial value of $\gamma$ resulted in consistent satisfaction of the constraint imposed on the QUBO in Equation \ref{eq:27}. Also, after aggregating the graph at the end of each inner while-loop (Algorithm \ref{alg:1}), $\gamma$ is updated to account for the increasing degrees of nodes:
\begin{equation} \label{eq:gammaUpdate}
         \gamma_{new} = max(k_i)\ \text{for}\ i \in G_{aggregated}
\end{equation} 

\subsubsection{Strategies for Reducing Runtime}
The majority of the different types of specialized hardware are mainly accessible via the cloud, commonly provided by some WebAPI access to the hardware. In order to make proposed approach applicable to a variety of specialized-hardware platforms, we take this into account and outline strategies that minimize the number of calls to the hardware, and if possible, minimize the size of the QUBO per iteration. 
\begin{enumerate}
    \item When set $S$ had only a single node, a greedy search replaced the call to the solver, and the node was assigned to the top choice determined by the greedy search. This strategies is especially helpful in reducing the number of calls to the solver when many nodes in the graph have been visited enough number of times by the algorithm, and hence the algorithm frequently selects single-member $S$ sets.
    
    \item When creating the candidate clusters set ($L$), nodes with only a single candidate cluster were excluded from set $S$, and instead were assigned to their single candidate clusters and were treated as normal nodes with determined clusters. After solving the QUBO for the reduced set $S$, the assignments for the single-candidate nodes were added back to the assignments determined by the solver. This strategy led to up to 50-60\% reductions in the size of the QUBOs sent to the solver for large QUBOs, and up to 30-40\% reduction in the number of calls made to the solver. This strategy yields more highlighted results after the first few passes over the graph in the inner while-loop (see Algorithm \ref{alg:1}), when the number of clusters in the graph shrinks significantly, and consequently the algorithm identifies many single-candidate nodes.
\end{enumerate}

\section{Results and Discussion}
\subsection{Experiments}
Table \ref{tab:1} lists the 12 benchmark graphs that are used in this paper: [Zachary's] Karate Club \cite{zachary1977information}, Meredith \cite{meredith1973regular}, Les Miserables \cite{knuth1993stanford}, and the rest of the graphs were selected from SNAP library \cite{snapnets}. The original Louvain algorithm was used to produce the baseline results of clustering, and Leiden algorithm \cite{traag2019louvain} (an updated version of Louvain) was used to produce rivaling results to compare the performance of the Ising-Louvain method against. To obtain the Louvain and Leiden results, each were run 20-30 times and the best results were selected. To obtain the Ising-Louvain results, several experiments were conducted to find the best combination of hyperparameters, and the results of the best experiments were reported. The Ising-Louvain algorithm was developed in Python 3.7 \cite{van2007python} using NetworkX v2.4 \cite{hagberg2008exploring} as the main library for handling the graphs, and the QUBOs were solved using Fujitsu's Digital Annealer.

In Table \ref{tab:1}, \#Clusters columns show the final number of clusters obtained by each method, Avg QUBO Size shows the average number of variables in QUBOs sent to the solver, and \#Solver-Calls shows the number of calls to the solver. 

\subsection{Results}
Table \ref{tab:1} shows that in all cases, Ising-Louvain has been able to obtain a more optimal clustering than that obtained by Louvain. The only exceptions are the three smallest graphs (Karate Club, Meredith, and Les Miserables) on which both Louvain and Ising-Louvain have obtained the globally optimal clustering; thus, the final modularities obtained by both methods are equal. This is largely because Ising-Louvain, using local optimization, can consider and evaluate orders of magnitude more possible assignments of clusters to nodes than Louvain is able to consider with a greedy approach. This way, Ising-Louvain can find more optimal clusterings than the original Louvain.


\begin{figure}
    \centering
    \includegraphics[width=0.48\textwidth]{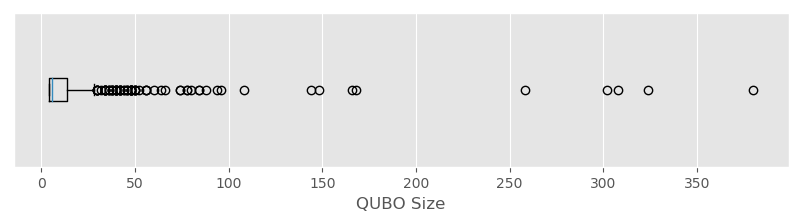}
    \caption{Boxplot of sizes of QUBOs solved for clustering the Facebook graph}
    \label{fig:3} 
\end{figure} 



For a more detailed analysis of QUBO sizes, Figure \ref{fig:3} shows the box-plot of the sizes of all QUBOs solved when clustering the Facebook graph (\textit{c.f.} Experiment 4 in Table \ref{tab:1}): the box starts from the first quantile (Q1 = 4) and continues until the third quantile (Q3 = 14) of the sizes with a blue line indicating the median value (Q2 = 6). The lower and upper whiskers correspond to, respectively, $Q1 - 1.5*(Q3-Q1)$ and $Q3 + 1.5*(Q3-Q1)$, and the circles show the data lying outside the boundary of the whiskers. Figure \ref{fig:3} shows that 75\% of the QUBOs sent to the solver had less than 14 variables, and most of the QUBOs had less than 50 variables. 

\begin{table}[t]
    \centering
    \caption{Benchmark instances}
    \label{tab:graphs}
    \begin{tabular}{lcccccc}
       \textbf{Graph Name} & KarateClub & Meredith & LesMiserables & Facebook & Autonomous & LastFM\\ \hline
       \textbf{$|V|$} & 34 & 70 & 77 & 4,039 & 6,474 & 7,624\\
       \textbf{$|E|$} & 78 & 140 & 254 & 88,234 & 13,895 & 27,806\\ \\
       \textbf{Graph Name} & arXiv & arXiv2 & AstroPh & Enron & DBLP & Amazon\\ \hline
       \textbf{$|V|$} & 9,877 & 12,008 & 18,772 & 36,692 & 317,080 & 334,863\\
       \textbf{$|E|$} & 25,998 & 118,521 & 198,110 & 183,831 & 1,049,866 & 925,872
    \end{tabular}
\end{table}


\begin{table}[t]
    \centering
    \caption{Summary of the results of running Louvain, Leiden, and Ising-Louvain on the benchmark graphs, along with statistics of the solver} 
\label{tab:1}
    \begin{tabular}{lccc|cc}
        \multicolumn{1}{l}{} & \multicolumn{3}{c}{\textbf{Optimal Modularity Results}} & \multicolumn{2}{c}{\textbf{Solver Stats}}\\
     \textbf{Graph} & \textbf{Leiden\ } & \textbf{Louvain\ } & \textbf{Ising-Louvain\ } & \textbf{\ Avg QUBO Size\ \ } & \textbf{\#Solver-Calls}\\
    \hline
    KarateClub & 0.4198 &\textbf{0.4198}& \textbf{ 0.4198 }& 75 & 4 \\
    Meredith & 0.7457 & \textbf{0.7571} & \textbf{0.7571} & 164  & 4 \\
    LesMiserables & 0.5600 & \textbf{0.5667} & \textbf{0.5667}& 85 & 14 \\
    Facebook & 0.8356 & 0.8350 & \textbf{0.8358} & 19 & 584 \\
    Autonomous & \textbf{0.6662} & 0.6553 & 0.6572 & 7 & 1626 \\
    LastFM &\textbf{0.8170} & 0.8155 & 0.8168 & 16 & 19036\\
    arXiv & \textbf{0.7759 } & 0.7699 & 0.7756 & 11 & 36373\\
    arXiv2 & 0.6649 & 0.6622 & \textbf{0.6650} & 32 & 115898\\
    AstroPh & 0.6367 & 0.6309 & \textbf{0.6378} & 44 & 351410\\
    Enron & 0.6265 & 0.6204 & \textbf{0.6285} & 23 & 146451\\
    DBLP & \textbf{0.8301} & 0.8220 & 0.8243 & 13 & 8369823\\
    Amazon & \textbf{0.9309}  & 0.9263 & 0.9277& 14 & 7812919\\
    \end{tabular} 
\end{table}
\section{Conclusion}
This paper introduces a new clustering algorithm based on the popular Louvain clustering algorithm for large-scale graphs using quadratic unconstrained binary optimization (QUBO). This method, called Ising-Louvain, is particularly fit for being used with specialized hardware, including Digital Annealers, which are capable of solving combinatorial optimization problems. The method builds around replacing with local optimization a greedy approach taken by most state-of-the-art clustering algorithms. The method proves effective in obtaining improved results (as measured by modularity metric) compared to Louvain and Leiden methods — the state-of-the-art in large-scale graph clustering — when tested on 12 graphs with sizes ranging from 34 to 335k nodes. The improvement is attributed to the ability of the proposed method in evaluating orders-of-magnitude more possibilities of assigning clusters to individual nodes, and hence being able to find more optimal clusterings than the greedy method can find. 


One noteworthy extension to this work can focus on using greedy approaches, like Leiden and Louvain, to run the initial iterations of the proposed algorithm on very large graphs, when the number of neighboring clusters for each node are large and hence the created QUBOs are massive. This can save a lot of computation time and can allow more experiments to be run on these graphs for obtaining more optimal solutions.

Future work can also try fitting the proposed algorithm with optimization objectives other than modularity, such as betweenness, to see the behavior of the algorithm with those objective functions. Further, the distance metric used in this paper was the adjacency matrix. Future research can try focusing on graphs in which other measures of distance between nodes, such as Euclidean distance, can be applied for selecting candidate nodes and clusters and forming the QUBOs. This can extend the application of Ising-Louvain to, especially, Geospatial Clustering.

Finally, the success of this paper in revising and improving Louvain algorithm shows promise for applying the same local-optimization-oriented approach to other unsupervised learning algorithms which depend on local search heuristics.



\bibliographystyle{splncs04}
\bibliography{splncs041}

\end{document}